\documentclass[%
 reprint,
 showpacs,preprintnumbers,
 amsmath,amssymb,
 aps,
 pre,
 floatfix,
]{revtex4-1}

\usepackage{graphicx}
\usepackage{dcolumn}
\usepackage{bm}
\usepackage{tikz}  

\begin{document}

\preprint{APS/123-QED}

\title{On the transport of interacting particles in a chain of cavities: \\ 
	Description through a modified Fick-Jacobs equation}

\author{G. P. Su\'arez}
\email{gsuarez@mdp.edu.ar}
\author{M. Hoyuelos}
\author{H. O. M\'artin}

\affiliation{Instituto de Investigaciones F\'isicas de Mar del Plata (IFIMAR -- CONICET)}
\affiliation{Departamento de F\'isica, Facultad de Ciencias Exactas y Naturales,
              Universidad Nacional de Mar del Plata.\\
              De\'an Funes 3350, 7600 Mar del Plata, Argentina}

\date{\today}

\begin{abstract}
We study the transport process of interacting Brownian particles in a tube of varying cross section. To describe this process we introduce a modified Fick-Jacobs equation, considering particles that interact through a hard-core potential. We were able to solve the equation with numerical methods for the case of symmetric and asymmetric cavities. We focused in the concentration of particles along the direction of the tube. We also preformed Monte Carlo simulations to evaluate the accuracy of the results, obtaining good agreement between theory and simulations.
\end{abstract}


\maketitle

\section{Introduction}

The diffusion of particles in a narrow channel is a subject that has brought attention in the last decade because of its ubiquity in nature. Examples of this can be found within blood vessels\cite{pries96,hedlishvili2001}, protein channels in cell membranes \cite{zhou2008}, zeolites \cite{karger, keil} and nano--porous materials \cite{beerdsen} among others. Theoretical description of these systems have been proposed using the Fick--Jacobs equation. For example, the problem of a Brownian particle diffusing in smoothly corrugated channels was studied in \cite{rubi2010, hanggi2009, burada2008, Ai2006}. Bruna \textit{et.al.} \cite{bruna} also investigated the case of finite sized  particles interacting through a hard--core potential and diffusing in a confined environment.

We will focus on identical particles, interacting through a hard-core potential. If the system is one-dimensional, the process is known as single file diffusion, see e.g. \cite{harris, richards, beijeren}. In this paper we are interested in diffusion along a narrow channel of variable transverse area and in the effects of a repulsive interaction among particles. We propose a generalization of the Fick-Jacobs equation in order to take into account the interaction. First we have to obtain the Fokker-Planck equation for interacting particles, since the Fick-Jacobs equation is derived from it. We arrive to the non-linear Fokker-Planck equation for Fermions and Bosons that can be reinterpreted for particles with repulsive or attractive interactions.

This paper is organized as follows. In Sect. \ref{sfj} we introduce the Fick-Jacobs equation. In Sect. \ref{sfp} we derive the non-linear Fokker-Planck equation for a given class of repulsive or attractive interaction potentials. In Sect. \ref{sfji} we use the expression for the particle current derived in Sect. \ref{sfp} to obtain a non-linear Fick-Jacobs equation that takes into account the hard-core interaction.
We test these novel Fick-Jacobs equation by solving it using the two cavities depicted in Sec.~\ref{sec:cavities} and compared this result with the one obtained from Monte Carlo simulations in Sec.~\ref{sec:numerical}. Some final remarks are stated in Sec.~\ref{sec:conclusions}.

\section{Fick-Jacobs equation}
\label{sfj}

Jacobs \cite{jacobs} introduced an effective one-dimensional diffusion equation, usually referred to as Fick-Jacobs equation, for the description of diffusion of non-interacting particles in a tube of varying cross section. If the cross section is constant, the problem is reduced to a simple one dimensional process. But if there is a region with, for example, a larger transverse area, a Brownian particle spends more time exploring the available space and the random walk along the direction of the tube is slowed.

Let us consider the $x$ axis along the center of the tube; $A(x)$ is the transverse area at $x$. For simplicity, we use two dimensional systems, in which $A(x)$ is the width of the channel (it is straightforward to extend the results to three dimensions). The local concentration is $c(x,y,t)$ and evolves according to the diffusion equation
\begin{equation*}
\frac{\partial c}{\partial t}= D\left( \frac{\partial^2 c}{\partial x^2}  + \frac{\partial^2 c}{\partial y^2} \right) 
\end{equation*}
where $D$ is a constant diffusion coefficient and the tube geometry is taken into account in the boundary conditions. Jacobs gives an equation for the reduced one dimensional concentration $g(x,t) = \int c(x,y,t) \,dy$:
\begin{equation}
\frac{\partial g}{\partial t} =  D \frac{\partial}{\partial x}\left( \frac{\partial g}{\partial x} - \frac{A'(x)}{A(x)}g \right)
\label{fj1}
\end{equation}
where $A'(x) = \frac{d A}{d x}$.
At equilibrium, $c$ is constant and $g$ is proportional to $A(x)$. An external potential $U(x)$ is included in the more general version \cite{zwanzig}:

\begin{equation}
  \begin{split}
  \frac{\partial g}{\partial t} = D \frac{\partial}{\partial x} e^{-\beta H(x)} \frac{\partial}{\partial x} e^{\beta H(x)}\,g  \\
  = D \frac{\partial}{\partial x}\left( \frac{\partial g}{\partial x} + \beta H'(x)\, g \right)
  \label{fj2}
  \end{split}
\end{equation}
where $H(x) = U(x) - \beta^{-1} \log A(x)$ is the free energy and $k \log A(x)$ is interpreted as the entropy ($\beta = 1/kT$). The main assumption in the derivation of (\ref{fj1}) or (\ref{fj2}) is local equilibrium, i.e., for each point $x$, the particle distribution in the transverse direction is approximately homogeneous, so that we can write
\begin{equation}
 c(x,y,t) \simeq \left\{ \begin{array}{ll}
	 g(x,t)/A(x) & \mbox{inside the channel} \\
	 0 & \mbox{outside}
 \end{array} \right.
\label{aprox}
\end{equation}

The aim is to derive a Fick-Jacobs equation for particles with hard core interaction. In the next section we obtain, as a first step, the diffusion equation for interacting particles. Then we obtain the corresponding Fick-Jacobs equation.

\section{Diffusion with interaction}
\label{sfp}

Let us consider a one dimensional lattice; $n_i^o=0$ or 1 is the occupation number of site $i$, and consecutive sites are separated a distance $a$ (as shown below, the results are easily generalized to higher dimensions). The current between sites $i$ and $i+1$ for a given configuration is:
\begin{equation}
J_i^o = n_i^o P_{i,i+1} - n_{i+1}^o P_{i+1,i}
\label{current}
\end{equation}
where $P_{i,i+1}$ is the transition rate from $i$ to $i+1$, that satisfies the relation
\begin{equation}
\frac{P_{i,i+1}}{P_{i+1,i}} = e^{-\beta(\Delta U_i + \Delta V_i)}.
\end{equation}
We have separated the external potential $\Delta U_i = U_{i+1}-U_i$ from the interaction potential $\Delta V_i = V_{i+1}-V_i$, that depends on the configuration $\{n_i\}$. It is convenient to define the transition rate
\begin{equation}
P = P_{i,i+1} e^{\beta (\Delta U_i + V_{i+1})} = P_{i+1,i} e^{\beta V_{i}},
\end{equation}
so that the current (\ref{current}) takes the form
\begin{equation}
J_i^o = n_i^o P e^{-\beta (\Delta U_i + \beta V_{i+1})} - n_{i+1}^o P e^{- \beta V_{i}}.
\end{equation}
The interaction potential in site $i$, $V_i$, is infinite if the site is occupied and 0 otherwise, so that $e^{-\beta V_i} = 1 - n_i^o$.
We assume that the variation of the external potential is small $\beta |\Delta U_i| \ll 1$, then
\begin{equation}
J_i^o = P (1-\beta \Delta U_i)\, n_i^o\, (1 - n_{i+1}^o) - P \, n_{i+1}^o\, (1 - n_i^o).
\label{curr2}
\end{equation}
We are interested in the configuration average of the current, $J_i = \langle J_i^o \rangle$, and the occupation number, $n_i = \langle n_i^o \rangle$. We take the decorrelation of the product $\langle n_i^o n_{i+1}^o\rangle \simeq n_i n_{i+1}$ (this does not usually hold for occupation number of tagged particles, as explained in \cite{suarez-sfd}, but in our case particles are indistinguishable and it turns out that the decorrelation is a valid approximation). In the continuous limit we replace the discrete variables by fields as follows: $n_i/a \rightarrow c(x)$ , $J_i \rightarrow J(x)$ and $U_i \rightarrow U(x)$, with $x=i a$; we approximate $\frac{n_{i+1}-n_i}{a^2} = \frac{\partial c}{\partial x}$ and consider that the concentration is smooth, $a^2 \frac{\partial c}{\partial x} \ll 1$. After some algebra, from (\ref{curr2}) we obtain
\begin{equation}
J = - D \left( \frac{\partial c}{\partial x} + \beta U' c\, (1-ac) \right)
\label{curr3}
\end{equation}
where the diffusion coefficient $D$ is equal to $P a^2$. In the following, we use the dimensionless space scale $x\rightarrow xa$, so that the lattice spacing is eliminated in (\ref{curr3}). In this case, $c(x) = n(x)$, where $n(x)$ is the continuous limit of $n_i$. We will use $n$ instead of $c$ in order to make clear that we use the dimensionless space scale. The corresponding Fokker-Planck equation, generalized to higher dimensions, is
\begin{equation}
\frac{\partial n}{\partial t} =  D \nabla \cdot \left[ \nabla n + \beta\, \nabla U\, n\, (1- n) \right]
\label{fpeq}
\end{equation}

The result can be generalized to a wider class of repulsive potentials: $e^{-\beta V_i} = 1- n_i^o/b$ ($b \in \mathbb{N}$). The potential becomes infinite when site $i$ is occupied by the maximum number of particles: $b$. The same equation (\ref{fpeq}) is obtained, the only difference is a scaling of the concentration $n \rightarrow n/b$. A value of the concentration equal 1 means that the maximum number of particles is reached.

A further generalization is to consider attractive potentials of the form $e^{-\beta V_i} = 1 +  n_i^o/b$. In this case, using again the scaling $n \rightarrow n/b$, the equation becomes
\begin{equation}
\frac{\partial n}{\partial t} =  D \nabla \cdot \left[ \nabla n + \beta\, \nabla U\, n\, (1+ n) \right]
\label{bosons}
\end{equation}

We arrived to the well known Fokker-Planck equations for Fermions (\ref{fpeq}) and Bosons (\ref{bosons}) \cite{frank} with the peculiarity of a classical context. It can be shown that the stationary solutions (zero current) correspond to the Fermi-Dirac and Bose-Einstein statistics.

\section{Fick-Jacobs equation with hard-core interaction}
\label{sfji}

We consider a two dimensional channel. The diffusion of particles takes place in this confined environment. We are looking for a reduced description in a one dimensional space (the coordinate along the direction of the channel), including effects of the hard-core interaction. We consider an external potential $U(x)$ independent of the transverse direction.

Integrating the continuity equation $\frac{\partial n}{\partial t} = - \nabla\cdot\mathbf{J}$ along the transverse direction $y$, we obtain
\begin{equation}
\frac{\partial g}{\partial t} = - \frac{\partial}{\partial x} A J
\end{equation}
where $g = \int dy \, n \simeq n A$ and we have approximated $\int dy\, J \simeq J A$. 
Since the concentration $n$ is approximately constant along the $y$ direction (local equilibrium assumption), then the $x$-current $J$, given by (\ref{curr3}), is also constant in the transverse direction. We obtain:
\begin{equation}
\frac{\partial g}{\partial t} =  D \frac{\partial}{\partial x} \left(\frac{\partial g}{\partial x} + \beta U' \,g\,\left(1 - \frac{g}{A}\right) - \frac{A'}{A} g \right)
\label{fjeq}
\end{equation}

The expression takes a more compact form in terms of $n$:
\begin{equation}
\frac{\partial n}{\partial t} = \frac{D}{A} \frac{\partial}{\partial x} A \left(\frac{\partial n}{\partial x} + \beta U' \,n\,\left(1 - n\right) \right)
\label{fjeqn}
\end{equation}
For the case of non-interacting particles, replacing $(1-n)$ by 1 in (\ref{fjeqn}), 
we recover the corresponding Fick-Jacobs equation (\ref{fj2}).

Let us consider the transversely integrated current $j=JA$ produced by $-U' = f/\beta$, where $f$ is a dimensionless force. The condition $\beta |\Delta U_i| \ll 1$ implies that $|f| \ll 1$. From Eq.~(\ref{fjeqn}) we have:
\begin{equation}
  A\frac{\partial n}{\partial x} - f A n (1-n) = -j/D
  \label{eq:fick-jacobs-final}
\end{equation}

We are interested in the stationary state with a constant force, were $j$ and $D$ are constants. In order to find the solution to this equation we use the expansion:
\begin{equation}
  n(x)=b_0+\sum_{k=1}^{\infty‎}a_k\:\sin\left(\frac{2\pi k}{L_x}x\right)+b_k\:\cos\left(\frac{2\pi k}{L_x}x\right),
  \label{eq:desarrollo}
\end{equation}
with the condition
\begin{equation}
  \int_0^{L_x}A(x)n(x)dx=N,
  \label{eq:m0}
\end{equation}
where $N$ is the total number of particles in the cavity, and $L_x$ is the length of the cavity. 

\section{Cavities}\label{sec:cavities}
\begin{figure}
  \begin{minipage}[b]{0.95\linewidth}
      \includegraphics[width=1\textwidth]{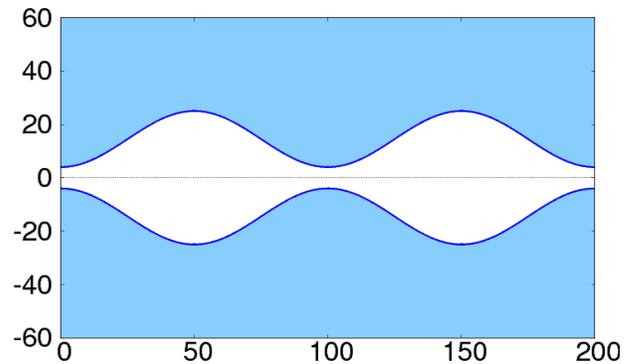}
      $(a)$
  \end{minipage}
  \vspace{0.5cm}
  \begin{minipage}[b]{0.95\linewidth}
      \includegraphics[width=1\textwidth]{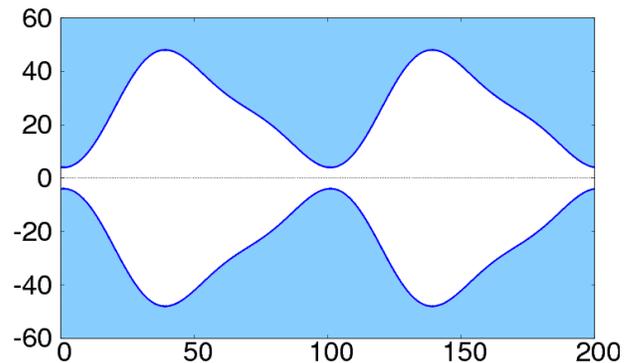}
      $(b)$
  \end{minipage}
  \caption{(color online) $(a)$ Symmetric and $(b)$ asymmetric cavities used for testing the theory. The distances are expressed in terms of the lattice spacing $a$. See the text for further details.}
  \label{fig:esquemas}
\end{figure}

To test its accuracy, Eq.~(\ref{eq:fick-jacobs-final}) has been solved using smooth cavities, as shown in Fig.\ref{fig:esquemas}. In behalf of simplicity, we selected cavities that are a linear combination of trigonometric functions. In addition to this, we created a discretized version of the same cavities depicted inside a square lattice. Monte Carlo simulations were carried out and we compared the solutions obtained from these two methods. Further details of the numerical simulations are included in Sec.~\ref{sec:numerical}.

Inside each of these cavities we are going to consider a mean concentration of particles, 
\begin{equation}
    \bar{c}=\frac{N}{\int_0^{L_x}A(x)dx}
\end{equation}
where, in the discrete case, $\int_0^{L_x}A(x)dx$ is equal to the total number of sites in the cavity.

In the symmetric case we use a cavity with a cosine shape, with top and bottom limits given by
\begin{equation}
  \begin{split}
      w_{\mathrm{top}} &= \alpha\left[1-\cos\left(\frac{2\pi}{L_x}x\right)\right] + \gamma \\ 
      w_{\mathrm{bottom}} &= -w_{\mathrm{top}} 
  \label{eq:sim-lat}
  \end{split} 
\end{equation}
so that the transverse width is $A(x) = 2w_{\mathrm{top}}$. 

We used the following set of parameters in order to characterize the symmetric cavity: $\alpha=10.5$, $\gamma=4$, $L_x=100$.
It is worth mentioning that in this case, if $\bar{c}=0.5$ only the coefficients $a_k$ in (\ref{eq:desarrollo}) are different from zero, i.e., $b_k=0\;\forall k=1,2,\cdots ,\infty$ and $b_0=\bar{c}$.

We also used Eq.~(\ref{eq:fick-jacobs-final}) to find the steady state of the transversally integrated concentration along the direction of the channel, $n(x)$, in an asymmetric cavity. This is, probably, the most interesting case because a different behavior can be obtained when the force is applied in opposite directions. It is specially compelling because of its application in devices to separate particles of different sizes \cite{Reguera2012, matthias}. The top and bottom borders of the asymmetric cavity shown in Fig.~\ref{fig:esquemas} are given by
\begin{equation}
  \begin{split}
  w_{\mathrm{top}} &= \alpha \left[\sin\left(\frac{2 \pi}{L_x} x\right) + \frac{\Delta}{4}\sin\left(\frac{4 \pi}{L_x} x\right)\right] + \gamma \\
  w_{\mathrm{bottom}}&=-w_{\mathrm{top}}
  \label{eq:asymm-lat} 
  \end{split}
\end{equation}
so that $A(x) = 2w_{\mathrm{top}}$. 

\begin{figure}
  \begin{minipage}[b]{0.8\linewidth}
      \includegraphics[width=1\textwidth]{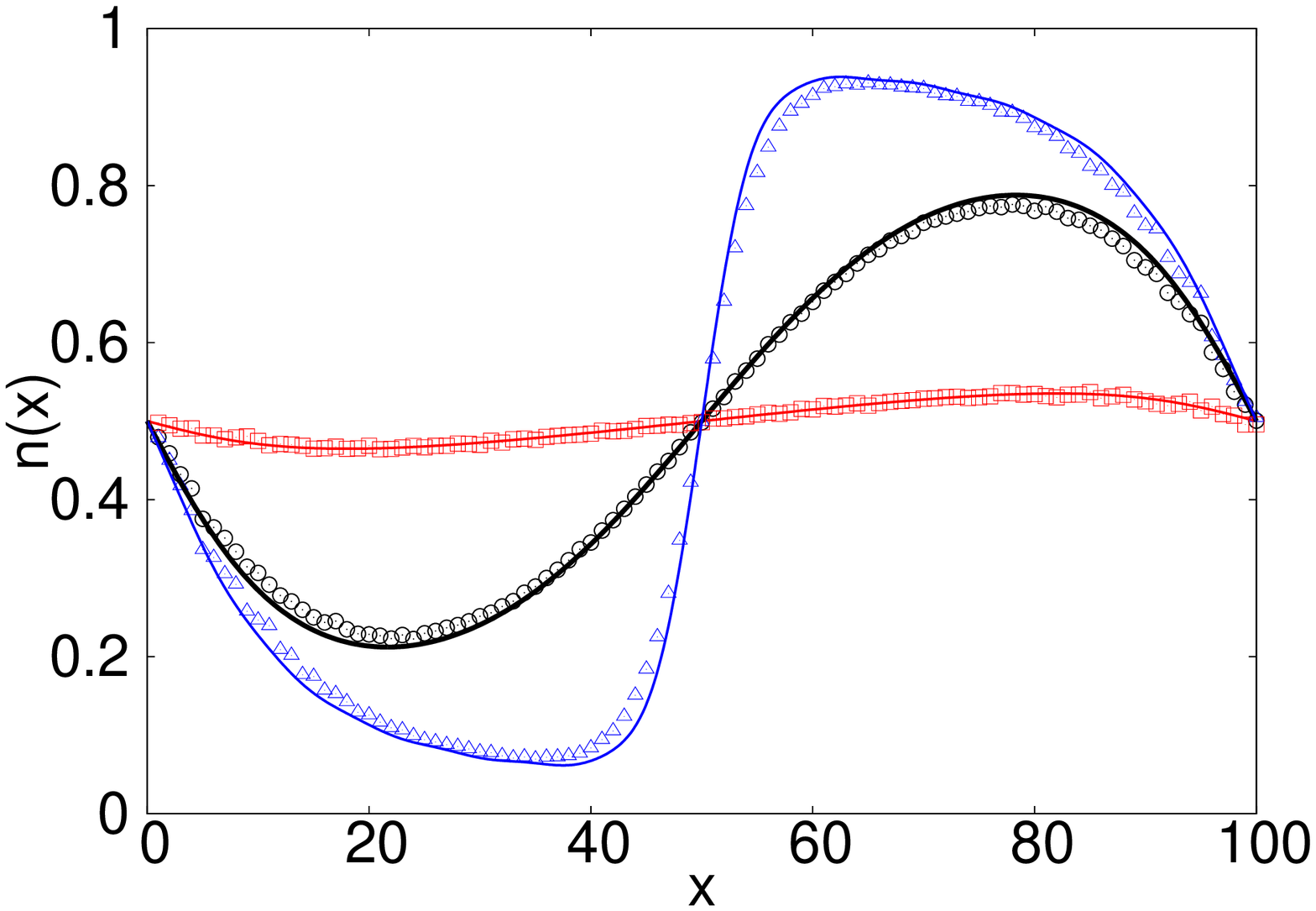}
      $(a)$
  \end{minipage}
  \hspace{0.5cm}
  \begin{minipage}[b]{0.8\linewidth}
      \includegraphics[width=1\textwidth]{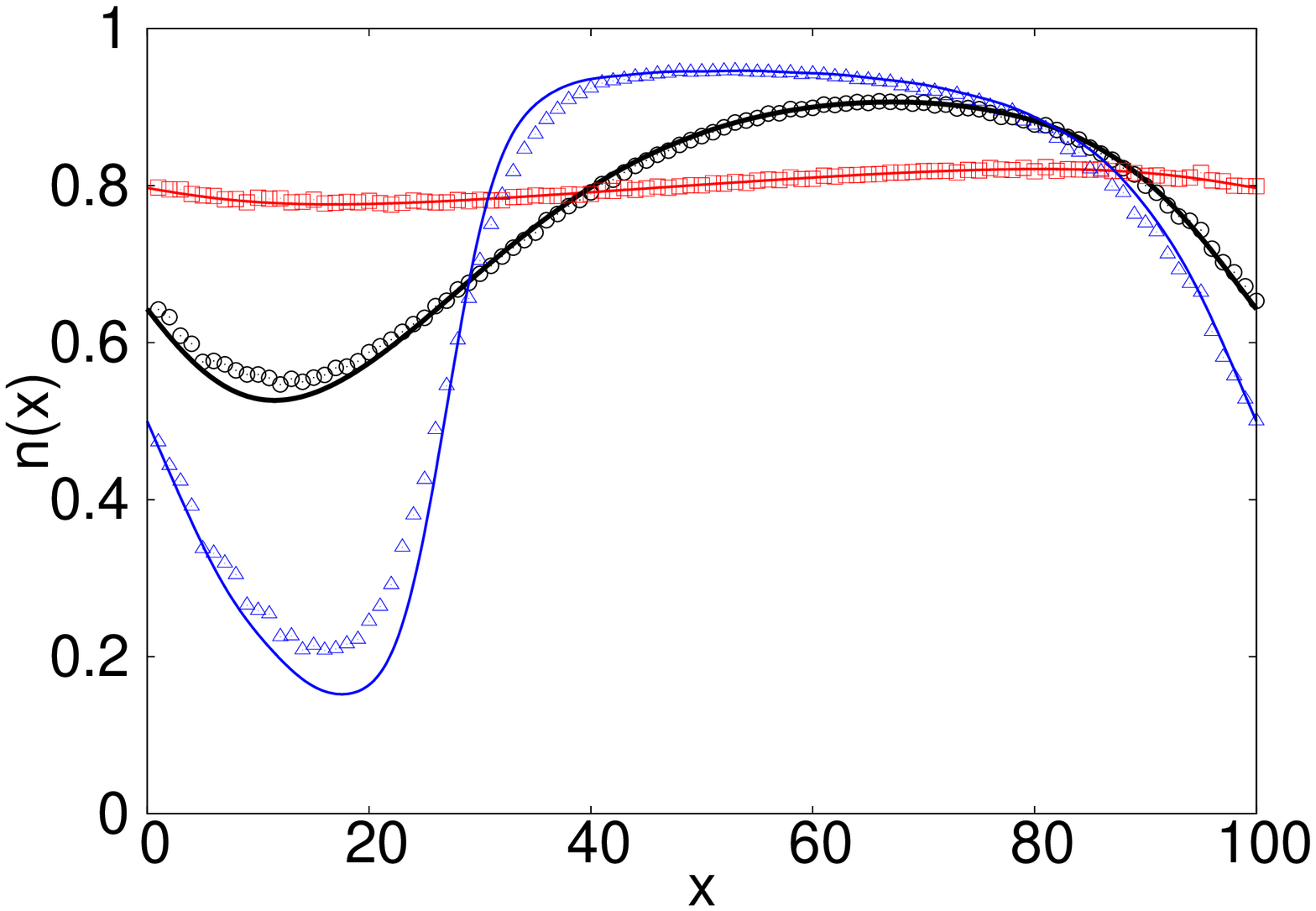}
      $(b)$
  \end{minipage}
  \caption{(color online) Occupation frequency in the stationary state against position $x$ in a symmetric lattice. $(a)$ $\bar{c}=0.5$, $(b)$ $\bar{c}=0.8$. For each mean concentration, the values of the force are $f=0.01$ (red squares), $0.1$ (black circles) and $0.5$ (blue triangles). Parameters of the cavity in (\ref{eq:sim-lat}): $\alpha=10.5$, $\gamma=4$, $L_x=100$. Full lines were obtained with $15$ terms of Eq.~(\ref{eq:desarrollo}). Dots were obtained with numerical simulations. Average over $10^6$ samples between $t=10^4$ and $t=10^8$ Monte Carlo time-steps.}
  \label{fig:simetrica-05-08}
\end{figure}

We used the following set of parameters to characterize the cavity: $\Delta=1$, $\alpha=20$, $\gamma=26$, $L_x=100$. Considering this values, and  $0\leq x\leq L_x$, Eq.~(\ref{eq:asymm-lat}) presents its minimum at $x\simeq 20$, so all the results have been shifted horizontally $20$ units in order to place the bottleneck in $x=0$ and $x=L_x$.
\section{Numerical Simulations} \label{sec:numerical}
\begin{figure}
  \begin{center}
    \includegraphics[width=0.7\linewidth]{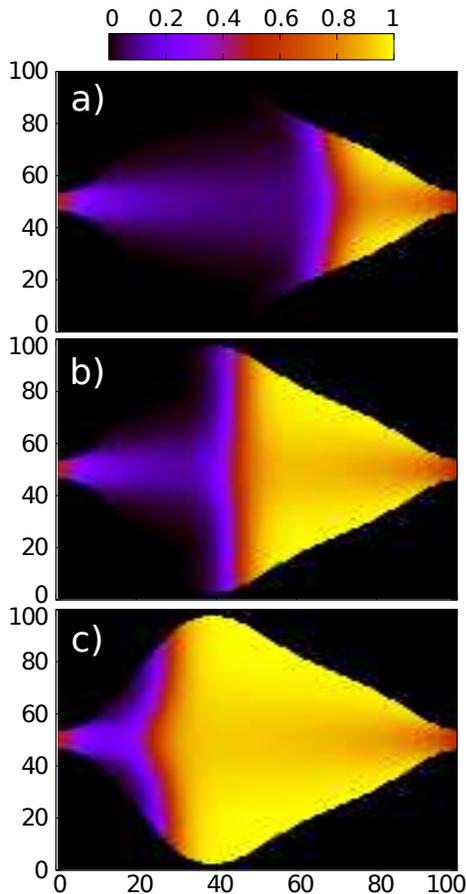}
    \caption{(color online) Numerical simulation of average density or site occupation frequency in an asymmetric cavity, in the stationary state. Values of the mean concentration: $(a)$ $\bar{c}=0.2$, $(b)$ $\bar{c}=0.5$, and $(c)$ $\bar{c}=0.8$. In all cases the force is $f=0.5$ (to the right). Number of samples: $10^5$.}
    \label{fig:escaladecolor}
  \end{center}
\end{figure}

 With the purpose of confirming the prediction of Eq.~(\ref{eq:fick-jacobs-final}), Monte Carlo simulations were carried out in cavities of different shapes. We analyzed two cases, symmetric and asymmetric cavities (Fig.~\ref{fig:esquemas}). In a square lattice of $L_x \times L_y$ sites, a discretized version of each cavity was depicted, indicating the region where the particles can diffuse. In this discretized version the contour of the cavities become irregular (see Fig.~\ref{fig:escaladecolor}). Within this region, $N$ particles are placed at random positions. They are allowed to jump freely inside the cavity, except when the new site is occupied with another particle. In this case the particle will remain on its site. We say that one Monte Carlo step has passed when, in average, every particle had had the opportunity to jump once. Periodic boundary conditions were set in the exits to simulate an infinite chain of cavities. 

\begin{figure}
  \begin{center}
    \includegraphics[width=0.7\linewidth]{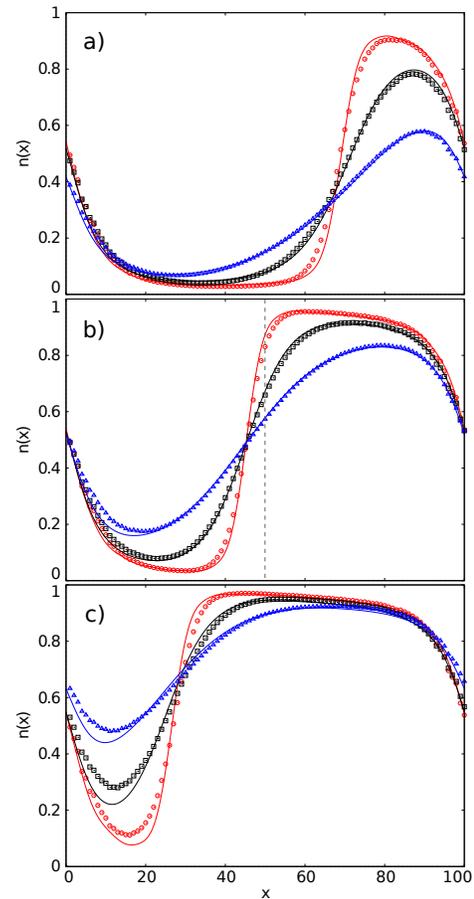}
    \caption{(color online) Stationary state of transversally integrated concentration along the direction of the channel in an asymmetric cavity. $(a)$ $\bar{c}=0.2$ $(b)$ $\bar{c}=0.5$ $(c)$ $\bar{c}=0.8$. In each case, the following values of the force were used: $f=0.1$ (blue triangles), $0.2$ (black squares) and $0.5$ (red circles). Solid lines were obtained with $12$ terms of Eq.~(\ref{eq:desarrollo}). Dots were obtained with simulations ($10^5$ samples between $t=10^4$ and $t=10^7$ Monte Carlo time-steps)}
    \label{fig:perfil-HC}
  \end{center}
\end{figure}

When there are no forces applied, all the particles have the same jumping rate in all directions. The jumps have a length equal to the size of a particle. If there is a force, $f$, acting along the $x$ axis, the jumping rate will be higher in the direction of the force. Thus, if the force is applied to the right, the jump rates in different directions are
\begin{equation}
  \begin{split}
    p_\uparrow=p_\downarrow=p_\leftarrow=p\\
    p_\rightarrow=p(1+f).
  \end{split}
  \label{eq:rate1}
\end{equation}

Likewise, if the force is applied to the left,
\begin{equation}
  \begin{split}
    p_\uparrow=p_\downarrow=p_\rightarrow=p\\
    p_\leftarrow=p(1+f).
  \end{split}
  \label{eq:rate2}
\end{equation}
We let the system evolve for some time, until it reaches the stationary state. After that we start to measure the amount of particles on every slice of the cavity of width $\delta x = 1$. After performing an average on configurations of this quantity, it is divided by the area of the cavity, $A(x)$, to find an estimation of $n(x)$.

In Figs.~\ref{fig:simetrica-05-08} $(a)$ and~$(b)$ we can see that there is a good agreement between analytical results (full lines) and numerical simulations (dots). For the symmetric cavity we show only the results with the force to the right, in view to the fact that the exact same behavior is expected when the force is applied in the opposite direction.

The results of the stationary concentration obtained with Monte Carlo simulations (dots) and with Eq.~(\ref{eq:fick-jacobs-final}) (full lines) are shown in Fig.~\ref{fig:perfil-HC} for the asymmetric cavity. As in the previous case, there is also a good agreement between theory and simulation. We use three values of the force and three different mean concentrations. In all cases the force is applied to the right. However, as a consequence of the particle-vacancy equivalence, it is easy to use the same data to obtain the results with the force to the left. As explained in \cite{suarez-cavidad} a system like this one with mean concentration $\bar{c}$ and a force magnitude $f>0$ (to the right) is completely analogous to a system with mean concentration $1-\bar{c}$ and force $-f$ (to the left), if we consider the diffusion of the holes instead of the particles. So we can write
\begin{equation}
  n_{f,\bar{c}}(x)=1-n_{-f,1-\bar{c}}(x)\:, 
\end{equation}
where subscripts indicate the corresponding values of force and mean concentration.

\begin{figure}
 \begin{center}
   \includegraphics[width=0.8\linewidth]{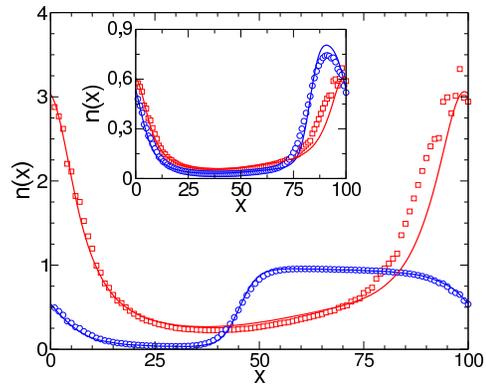}
   \caption{(color online) Transversally integrated concentration along the direction of the channel for interacting (blue circles) and non-interacting (red squares) particles in the asymmetric cavity. Solid lines were obtained with 12 terms of Eq.~(\ref{eq:desarrollo}). Dots were obtained with numerical simulations. Parameters: $\bar{c}=0.5$ -- $f=0.5$. Inset: same information but $\bar{c}=0.1$. In all cases, numerical simulations were averaged over $10^6$ samples between $t=10^4$ and $t=10^8$ Monte Carlo time-steps.}
   \label{fig:con-sin-HC}
 \end{center}
\end{figure}

The equivalence is based on the fact that for every particle that moves to the left there is an empty site that moves to the right, and vice versa. Let us stress that this symmetry holds in Eq.~(\ref{eq:fick-jacobs-final}) when $n\rightarrow 1-n$, $f\rightarrow -f$ and $j\rightarrow -j$.

In Fig.~\ref{fig:perfil-HC}b a dashed line corresponding to $x=50$ is shown. If the lattice were symmetric then, those three curves should be symmetric around the $x=50$ line. The fact that it is not, shows the effect of the asymmetry of the cavity in the transversally integrated concentration along the direction of the channel.

In Fig.~\ref{fig:escaladecolor} we can see the average site occupation frequency on each site. This gives us extra details about the distribution of particles inside the cavity. In the derivation of Eq.~(\ref{fjeqn}) it was assumed that $n$ is approximately constant along the transverse direction of the channel. It is important to check the accuracy of this assumption. As can be seen from Fig.~\ref{fig:escaladecolor}, the approximation is quite good for $\bar{c}=0.5$, but a deviation from a constant transverse concentration is visible for $\bar{c}=0.2$ or $\bar{c}=0.8$, specially near the exits. This explains the discrepancy between theory and simulation in, for example, Fig. \ref{fig:perfil-HC}(c) for $x\simeq 15$.
Note that we are dealing with a rather large value of the force ($f=0.5$), where the condition $|f|\ll 1$ does not hold (see above Eq.~(\ref{eq:fick-jacobs-final})).
Nevertheless, the equation derived is robust enough to still provide acceptable results in a considerable range of concentrations and values of the force, as shown in Figs.~\ref{fig:simetrica-05-08} and~\ref{fig:perfil-HC}. 

To highlight the effect of the hard-core interaction we obtain $n(x)$ for the same cavity and mean concentration considering interacting and non-interacting particles (see Fig.~\ref{fig:con-sin-HC}). The difference in the concentration profile is clearly visible specially in the region with large concentration.
A non-interacting system allows for multiple-occupation of sites. This gives a whole different dynamic to the problem because the accumulation of particles in the bottleneck does not affect their mobility. Also, in this case $n(x)$ is no longer restricted to the $[0;1]$ interval.
As expected the difference between interacting and non-interacting cases increase when the mean concentration of particles increase.

\section{Conclusions}
\label{sec:conclusions}
We obtained the Fick-Jacobs equation for the diffusion of interacting particles in a narrow channel with variable width. The particles are drifted by an external force $f$. The equation gives an approximation for the transversally integrated concentration that holds for $|f| \ll 1$. The accuracy degree of the approximation was checked with numerical simulations.

We obtained a good agreement between theoretical and numerical results even for rather large values of the force ($f=0.5$). This equation has been solved for the case of a symmetric and an asymmetric chain of cavities. The former was selected mainly because of the simplicity to describe it, and the latter because of the importance of the asymmetric diffusion process and its applications. These two geometries exemplify the accuracy of the equation obtained, but we are confident that similar results will be found with different cavities.

Finally, let us note that, using a discrete model, we derived, on one hand, the Fick-Jacobs equation for interacting particles and, on the other hand, the non-linear Fokker-Planck equation for Fermions and Bosons. 

\begin{acknowledgments}
This work was partially supported by Consejo Nacional de Investigaciones Científicas y 
Técnicas (CONICET, Argentina, PIP 0041 2010-2012).
\end{acknowledgments}

\end{document}